\documentclass[10pt,conference]{IEEEtran}
\usepackage{graphicx}
\usepackage{amsmath}
\usepackage{amssymb}
\usepackage[dvipdfm]{hyperref}
\usepackage{theorem}

\newif\ifdraft
\newif\iffigures

\drafttrue
\figurestrue

\graphicspath{{./}{/home/jungp/TeX/Figures/}}

\theoremstyle{plain}
\newtheorem{mytheorem}{Theorem}
\newtheorem{mycorollary}[mytheorem]{Corollary}

\theoremstyle{definition}

\newtheorem{mydefinition}[mytheorem]{Definition}

\usepackage{pifont}
\usepackage{graphics,graphicx,color}

\catcode`\@=11
\@ifpackageloaded{beamer}{
}{
}
\catcode`\@=12

\newcommand{\Trace}[1]{\,\mathbf{Tr}{#1}}

\newcommand{\defeq}{\overset{\text{def}}{=}}

\newcommand{\Pro}{\Pi}

\newcommand{\BHScat}{\boldsymbol{C}}

\newcommand{\Leb}[1]{\mathcal{L}_{#1}}
\newcommand{\Ltwo}{\Leb{2}}

\newcommand{\Wigner}{{\mathbf{W}}}
\newcommand{\Amb}{{\mathbf{A}}}

\newcommand{\Real}[1]{{\text{Re}}\{#1\}}

\newcommand{\Shift}{{\boldsymbol{S}}}

\newcommand{\esup}{\text{\rm ess sup\,\,}}

\newcommand{\Reals}{\mathbb{R}}
\newcommand{\RealsPlus}{\mathbb{R}_{+}}
\newcommand{\Complexes}{\mathbb{C}}

\newcommand{\Maximize}{\text{maximize }}

\begin{document}
\title{Weighted Norms of Ambiguity Functions and Wigner Distributions }
\author{Peter Jung\\
  Fraunhofer German-Sino Lab for Mobile Communications (MCI) and the Heinrich-Hertz Institute\\[.1em]
  \small{jung@hhi.fraunhofer.de}}
\maketitle
\begin{abstract}
   In this article new bounds on weighted $p$-norms of ambiguity functions and Wigner functions are 
   derived. Such norms occur frequently in several areas of physics and engineering. 
   In pulse optimization for Weyl--Heisenberg signaling in wide-sense stationary uncorrelated scattering 
   channels for example it is a key step to find the optimal waveforms for a given scattering statistics
   which is a problem also well known in radar and sonar waveform optimizations.
   The same situation arises in quantum information processing and optical communication when optimizing 
   pure quantum states for communicating in bosonic quantum channels, i.e. find optimal channel input states
   maximizing the pure state channel fidelity. 
   Due to the non-convex nature of this problem the optimum and the maximizers itself are in general difficult find, 
   numerically and analytically. Therefore upper bounds on the achievable 
   performance are important which will be provided  by this contribution. 
   Based on a result due to E. Lieb \cite{lieb:ambbound}, the main theorem
   states a new upper bound which is independent of the waveforms and becomes tight only for Gaussian weights
   and waveforms. A discussion of this particular important case, 
   which tighten recent results on Gaussian quantum fidelity and coherent states, will be given.
   Another bound is presented for the case where scattering is determined only by some arbitrary region
   in phase space.
\end{abstract}
\section{Introduction}
Time-frequency representations are an important tool in signal analysis, physics and many other scientific
areas. Among them are the Woodward cross ambiguity function $\tilde{\Amb}_{g\gamma}(\tau,\nu)$, 
which can be defined as ($\overline{\cdot}$ denotes complex conjugate)
\begin{equation}
   \tilde{\Amb}_{g\gamma}(\tau,\nu)=\int g(t-\frac{\tau}{2})\overline{\gamma}(t+\frac{\tau}{2})e^{-i2\pi\nu t}dt
\end{equation}
and the Wigner distribution $\Wigner_{g\gamma}(\tau,\nu)$
\begin{equation}
   \Wigner_{g\gamma}(\tau,\nu)=\int g(\tau+\frac{t}{2})\overline{\gamma}(\tau-\frac{t}{2})e^{-i2\pi\nu t}dt
\end{equation}
where the functions $g,\gamma:\Reals\rightarrow\Complexes$ assumed to be in 
$\Ltwo(\Reals)$\footnote{Which can be relaxed to other spaces by the H\"older inequality}. 
Both are related by $\Wigner_{g\gamma}(\tau,\nu)=2\tilde{\Amb}_{g\gamma^-}(2\tau,2\nu)$ 
where $\gamma^-(t)=\gamma(-t)$. Hence all results which will presented later on apply on 
Wigner functions as well.
Due to non--commutativity of the shifts in $\tau$ and $\nu$ (in phase space) there exists many definitions 
of these functions which differ only by phase factors. In considering norms only, the
ambiguities due to these phase factors are not important.

To be consistent with the previous work in \cite{jung:wssuspulseshaping,jung:isit05} 
in this article the alternative definition 
\begin{equation}
   \Amb_{g\gamma}(x)\defeq\langle g,\Shift_x\gamma\rangle=\int \overline{g}(t)(\Shift_x \gamma)(t)dt
\end{equation}
is used, where $\Shift_x$ is the time-frequency shift operator given as
\begin{equation}
   (\Shift_x f)(t)\defeq e^{i2\pi x_2 t}f(t-x_1)
\end{equation}
and $x=(x_1,x_2)\in\Reals^2$. 

Note that $\Amb_{g\gamma}(x)=e^{i\pi x_1 x_2}\tilde{\Amb}_{\overline{g}\overline{\gamma}}(-x_1,-x_2)$.
These operators establish up to phase factors an unitary representation of the Weyl--Heisenberg
group on $\Ltwo(\Reals)$ --- the so called Schr\"odinger representation 
(see for example \cite{folland:harmonics:phasespace}). 
They equal (again up to phase factors) the so called Weyl operators 
(Glauber displacement operators), i.e. perform phase space displacements in one dimension.

It is an important and in general unsolved (non--convex) problem in many fields of physics and engineering 
to find normalized function $g$ and $\gamma$ such that the following integral 
\begin{equation}
   \int |\langle g,\Shift_{x}\gamma\rangle|^2 \BHScat(x) dx
   =\int |\Amb_{g\gamma}(x)|^2 \BHScat(x) dx
   \label{eq:ambbound:problemintegral}
\end{equation}
is maximized where $\BHScat(x)$ could be some probability distribution and $dx$ is the Lebesgue
measure on $\Reals^2$. For example in radar and sonar
application (\ref{eq:ambbound:problemintegral}) is typically related to
the correlation response with some filter $g$ of a 
transmitted pulse $\gamma$ after passing through
a non-stationary scattering environment characterized by some $\BHScat(\cdot)$. This formulation
is obtained for so called Weyl--Heisenberg signaling in wide-sense stationary uncorrelated 
scattering (WSSUS) channels \cite{kozek:thesis,jung:wssuspulseshaping,liu:orthogonalstf,kozek:nofdm1} 
where $\BHScat(\cdot)$ is called the
scattering function. 

If considering $\gamma$ as
a probability wave function in quantum mechanics (\ref{eq:ambbound:problemintegral}) can be considered 
also as its overlap with some wave function $g$ after several phase space interactions.
In quantum information processing (\ref{eq:ambbound:problemintegral}) is typically written as
pure state fidelity
\begin{equation}
   (\ref{eq:ambbound:problemintegral})
   =\Trace\{\Pro_g \int \Shift_{x}\Pro_\gamma \Shift_{x}^* \BHScat(x)dx\}
   \defeq\Trace\{\Pro_g A(\Pro_\gamma)\}
   \label{eq:quantum:fidelity}
\end{equation}
where $\Pro_f\defeq\lVert f\rVert_2^{-2}\langle f,\cdot\rangle f$ is the rank-one projector 
onto $f$ and $\Trace(\cdot)$ denotes the trace functional. 
The middle term  in \eqref{eq:quantum:fidelity} is the Kraus representation \cite{kraus:states:effects}
of a bosonic quantum channel $A(\cdot)$ \cite{holevo:propquantum,hall:gaussiannoise} which maps
the input state $\Pro_\gamma$ (rank--one density operator) to the output $A(\Pro_\gamma)$.
Minimizing the probability of error 
$P_e=1-\Trace\{\Pro_g A(\Pro_\gamma)\}$ (see for example \cite{helstrom:quantumdet}) 
for rank--one measurements is then the maximization of the pure state fidelity, i.e.
the  following optimization problem:
\begin{equation}
   \underset{g,\gamma}{\Maximize}\Trace\{\Pro_g A(\Pro_\gamma)\}
   \label{eq:ambbound:traceform}
\end{equation}
For each $\gamma$ the operator $A(\Pro_\gamma)$ is a positive semi--definite trace class
(thus compact) operator, hence \eqref{eq:ambbound:traceform} likewise reads
\begin{equation}
   \underset{{\Trace{X}=1,X\geq 0}}{\Maximize}\lambda_{\max}(A(X))
   \label{eq:ambbound:outputpurity}
\end{equation}
where the rank--relaxation 
follows from convexity of the maximal eigenvalue $\lambda_{\max}(\cdot)$ and 
linear--convexity of $A(\cdot)$ (see for example \cite{jung:wssuspulseshaping,jung:isit05}).

In a slightly more general context this paper considers 
$\lVert|\Amb_{g\gamma}|^r\BHScat\rVert_1$ which directly gives
the weighted $r$--norms of ambiguity functions in the form of
\begin{equation}
   \lVert\Amb_{g\gamma}\rVert_{r,\BHScat}=\left(\int |\Amb_{g\gamma}(x)|^r \BHScat(x) dx\right)^{1/r}
   =\lVert|\Amb_{g\gamma}|^r\BHScat\rVert_1^{1/r}
\end{equation}
where $\BHScat:\Reals^2\rightarrow\RealsPlus$ is now some arbitrary weight function. For $r=2$ the results match then the 
examples given so far. Note furthermore that this topic is also connected
to R\'{e}nyi entropies $H(r)$ of time--frequency representations
\begin{equation}
   H(r)=\frac{1}{1-r}\log{\lVert\Amb_{g\gamma}\BHScat^\frac{1}{r}\rVert_r^r}
\end{equation}
i.e. a measure of time--frequency information content \cite{baraniuk:renyi}.

\section{Main Results}

The results are organized in a main theorem presenting the general upper bound to
$\lVert|\Amb_{g\gamma}|^r\BHScat\rVert_1$. Then, two special cases
are investigated in more detail. The first is dedicated to the overall equality case in the main theorem
and important for Gaussian bosonic quantum channels.
The second case discusses another application relevant situation motivated by
WSSUS pulse shaping in wireless communications.
But before starting, the following definitions are needed.

\begin{mydefinition}
   Let $0<p<\infty$. For functions $f:\Reals\rightarrow\Complexes$ and 
   $F:\Reals^2\rightarrow\Complexes$
   \begin{equation*}
      \lVert f\rVert_p\defeq\left(\int|f(t)|^pdt\right)^{1/p}\quad
      \lVert F\rVert_p\defeq\left(\int|F(x)|^pdx\right)^{1/p}
   \end{equation*}
   are then the common notion of $p$--norms,
   where $dt$ and $dx$ are the Lebesgue measure on $\Reals$ and $\Reals^2$.
   Furthermore for $p=\infty$ is
   \begin{equation*}
      \lVert f\rVert_\infty\defeq\esup |f(t)|\quad      \lVert F\rVert_\infty\defeq\esup |F(x)|
   \end{equation*}
   If $\lVert f\rVert_p$ is finite $f$ is said to be in $\Leb{p}(\Reals)$ (similarly 
   if $\lVert F\rVert_p$ is finite, $F$ is said to be in $\Leb{p}(\Reals^2)$).
\end{mydefinition}
For discussion of the equality case for the presented bound
the formulation ''to be Gaussian'' for functions $f:\Reals\rightarrow\Complexes$ and 
$F:\Reals^2\rightarrow\Complexes$ is needed.
\begin{mydefinition}
   Functions  $f(t)$ and $F(x)$ are said to be ''Gaussian'' if for
   $a,b,c,C\in\Complexes$, $A\in\Complexes^{2\times2}$ and $B\in\Complexes^{2}$
   \begin{equation}
      f(t)=e^{-at^2+bt+c}\quad F(x)=e^{-\langle x,Ax\rangle+\langle B,x\rangle+C}
   \end{equation}
   and $\Real{a}>0$ and $A^*A>0$.
   \label{def:gaussians}
\end{mydefinition}
Two Gaussians $f(t)$ and $g(t)$ are called {\it matched} if they have the same parameter $a$.
 
The main ingredient for the presented analysis is the following theorem due to E. Lieb \cite{lieb:ambbound} on
(unweighted) norms of ambiguity functions.  
\begin{mytheorem} (E. Lieb) 
   Let $\Amb_{g\gamma}(x)=\langle g,\Shift_{x}\gamma\rangle$ be the cross ambiguity function between 
   functions $g\in\Leb{a}(\Reals)$ and $\gamma\in\Leb{b}(\Reals)$ where $1=\frac{1}{a}+\frac{1}{b}$.
   If $2<p<\infty$ with $q=\frac{p}{p-1}\leq a\leq p$ and
   $q\leq b\leq p$, then holds
   \begin{equation}
      \lVert\Amb_{g\gamma}\rVert^p_p\leq H(p,a,b)\lVert g\rVert_a^p\lVert\gamma\rVert_b^p
      \label{eq:lieb:ambleq}
   \end{equation}
   where $H(p,a,b)=c^p_q\left(c_{a/q}c_{b/q}c_{p/q}\right)^{p/q}$, $c_p=p^{1/(2p)}q^{-1/(2q)}$. 
   Equality is achieved with $g$ and $\gamma$ being Gaussian if and only if both $a$ and $b>p/(p-1)$.
   In particular for $a=b=2$
   \begin{equation}
      \lVert\Amb_{g\gamma}\rVert^p_p\leq\frac{2}{p}\lVert g\rVert_2^p\lVert\gamma\rVert_2^p
      \label{eq:lieb:ambleqtwo}
   \end{equation}
   \label{thm:lieb:ambleq}
\end{mytheorem}
Actually Lieb proved also the reversed inequality for $1\leq p<2$. Furthermore, for the 
case $p=2$ it is well known that equality holds in (\ref{eq:lieb:ambleqtwo}) for all 
$g$ and $\gamma$. Then the optimal slope (related to entropy)
\begin{equation}
   \frac{1}{p}\int|\Amb_{g\gamma}(x)|^p\ln|\Amb_{g\gamma}(x)|^p dx
\end{equation}
at $p=2$ is achieved by matched Gaussians \cite{lieb:ambbound}. For simplifications it is assumed 
from now that $\lVert g\rVert_2=\lVert\gamma\rVert_2=1$.
With the previous preparations the main theorem in this article is now:
\begin{mytheorem} 
   Let $\Amb_{g\gamma}(x)=\langle g,\Shift_{x}\gamma\rangle$ be the cross ambiguity function between 
   functions $g,\gamma$ with $\lVert g\rVert_2=\lVert\gamma\rVert_2=1$ and $p,r\in\Reals$. Furthermore
   let $\BHScat(\cdot)\in\Leb{q}(\Reals^2)$ with $q=\frac{p}{p-1}$. Then 
   \begin{equation}
      \lVert |\Amb_{g\gamma}|^r \BHScat \rVert_1 \leq \left(\frac{2}{rp}\right)^{\frac{1}{p}}\lVert\BHScat\rVert_{\frac{p}{p-1}}
   \end{equation}
   holds for each $p\geq\max\{1,\frac{2}{r}\}$.
   \label{thm:jung:fidelitybound}
\end{mytheorem}
\begin{proof}
   In the first step H\"older's inequality gives
   \begin{equation}
      \lVert |\Amb_{g\gamma}|^r \BHScat \rVert_1 \leq \lVert|\Amb_{g\gamma}|^r\rVert_p \rVert\BHScat \rVert_q
      \label{eq:hoelderapplied}
   \end{equation}
   for conjugated indices $p$ and $q$, thus with $1=\frac{1}{p}+\frac{1}{q}$. 
   Equality is achieved for $1<p<\infty$ if and only if there exists $\lambda\in\Reals$ such that
   \begin{equation}
      |\BHScat(x)|=\lambda|\Amb_{g\gamma}(x)|^{r(p-1)}
      \label{eq:hoelderequality:conditions}
   \end{equation}
   for almost every $x$. Similar conclusions for $p=1$ and $p=\infty$ are not considered 
   in this paper.
   Lieb's inequality in the form of (\ref{eq:lieb:ambleqtwo}) for 
   $\lVert \Amb_{g\gamma}\rVert^{rp}_{rp}$ gives for 
   rhs of (\ref{eq:hoelderapplied}) 
   \begin{equation}
      \begin{split}
         \lVert|\Amb_{g\gamma}|^r\rVert_p \rVert\BHScat \rVert_q
         &=\lVert \Amb_{g\gamma}\rVert^r_{rp} \rVert\BHScat \rVert_q\\
         &=\left(\lVert \Amb_{g\gamma}\rVert^{rp}_{rp}\right)^{\frac{1}{p}} \rVert\BHScat \rVert_q         
         \leq\left(\frac{2}{rp}\right)^{\frac{1}{p}}\lVert\BHScat\rVert_q
      \end{split}
      \label{eq:liebapplied}
   \end{equation}
   The latter holds for every $rp\geq2$, thus the case $rp=2$ is now included as already mentioned before.
   Equality in (\ref{eq:liebapplied}) is achieved if $g$ and $\gamma$ are matched Gaussians. Furthermore if
   strictly $rp>2$, equality in (\ref{eq:liebapplied}) 
   is {\it only} achieved if $g$ and $\gamma$ are matched Gaussians.
   Replacing $q=\frac{p}{p-1}$ gives the desired result.      
\end{proof}
\vspace*{.5em}

Note that apart from the normalization constraint the bound in Thm.\ref{thm:jung:fidelitybound} 
does not depend anymore on $g$ and $\gamma$. Hence 
for any given  $\BHScat(\cdot)$ the optimal bound 
can be found by
\begin{equation}
   \min_{\Reals\ni p\geq\max\{1,\frac{2}{r}\}}\left(
     \left(\frac{2}{rp}\right)^{\frac{1}{p}}\lVert\BHScat\rVert_{\frac{p}{p-1}}
   \right)
\end{equation}
In the minimization $p\geq 1$ has to be forced for H\"older's inequality and 
$p\geq\frac{2}{r}$ for Lieb's inequality.
Two special cases are investigated now in more detail which are 
relevant for application.

First the overall equality case in Thm.\ref{thm:jung:fidelitybound}
is considered.
\begin{mycorollary}
   Let $\BHScat(x)=\alpha e^{-\alpha\pi (x_1^2+x_2^2)}$ with $\Reals\ni\alpha>0$. Then for each 
   $p\geq\max\{1,\frac{2}{r}\}$ holds
   \begin{equation}
      \lVert |\Amb_{g\gamma}|^r \BHScat \rVert_1 \leq 
      \left(\frac{2\alpha}{rp}\right)^{\frac{1}{p}}\left(\frac{p-1}{p}\right)^\frac{p-1}{p}
   \end{equation}
   The best bound is given as 
   \begin{equation}
      \lVert |\Amb_{g\gamma}|^r \BHScat \rVert_1 \leq 
      \begin{cases}
         \frac{2\alpha}{2\alpha+r} & \alpha\geq\frac{2-r}{2}\\
         \alpha^\frac{r}{2}(1-r/2)^{1-r/2}         & \text{else}
      \end{cases}
      \label{eq:ambbound:gaus:best}
   \end{equation}
   For $\alpha\geq\frac{2-r}{2}$ equality is achieved at $p=\frac{2\alpha}{r}+1$ 
   if and only if $g$ and $\gamma$ are matched Gaussian, i.e. then
   \begin{equation}
      \lVert |\Amb_{g\gamma}|^r \BHScat \rVert_1
      =\frac{2\alpha}{2\alpha+r}
   \end{equation}
   holds.
   \label{cor:ambbound:gaussian}
\end{mycorollary}

\begin{figure}[h]
   \includegraphics[width=\linewidth]{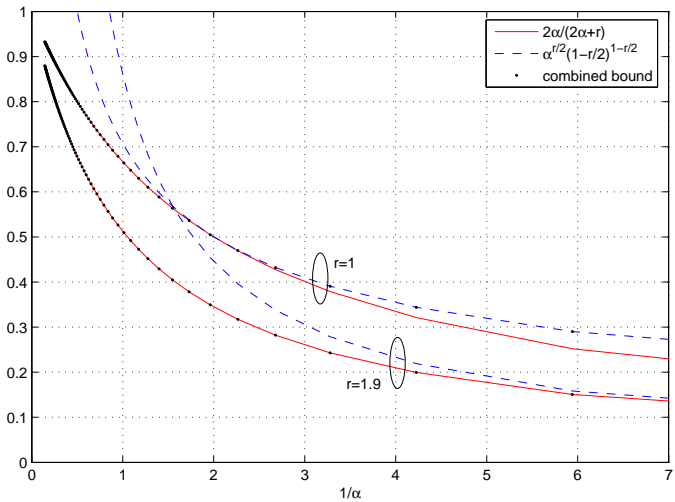}
   \caption{{\it Norm bounds for Gaussian weights:} Both functions in (\ref{eq:ambbound:gaus:best})
     separately and the combined version are shown for $r=1$ and $r=1.9$.}
   \label{fig:ambbound:gaus}
\end{figure}

\begin{proof}
   The moments of $\Leb{1}$--normalized two--dimensional Gaussians 
   are given as 
   \begin{equation}
      \lVert\BHScat\rVert_s=\left(\frac{1}{s}\right)^{\frac{1}{s}}\alpha^{\frac{s-1}{s}}
   \end{equation}
   According to Thm.\ref{thm:jung:fidelitybound} the upper bound 
   \begin{equation}
      \begin{split}
         f(p)
         &\defeq\left(\frac{2}{rp}\right)^{\frac{1}{p}}\lVert\BHScat\rVert_{\frac{p}{p-1}}
         ={\left(\frac{2\alpha}{rp}\right)}^{\frac{1}{p}}{\left(\frac{p-1}{p}\right)}^{\frac{p-1}{p}}
      \end{split}
   \end{equation}
   holds for each $p\geq\max\{1,2/r\}$.
   The optimal (minimal) bound is attained as some point $p_{\min}$ which can be obtained as
   \begin{equation}
      \min_{\Reals\ni p\geq\max\{1,\frac{2}{r}\}}f(p)=f(p_{\min})
   \end{equation}
   The first derivative $f'$ of $f$ at point $p$ is
   \begin{equation}
      f'(p)=\frac{f(p)}{p^2}\ln(\frac{r(p-1)}{2\alpha})
   \end{equation}
   Thus $f'(p_{\min})=0$ gives only one stationary point $p_{\min}$
   \begin{equation}
      \frac{r(p_{\min}-1)}{2\alpha}=1\quad\Leftrightarrow\quad p_{\min}=\frac{2\alpha}{r}+1> 1
   \end{equation}
   Due to $f(p)/p^2>0$ and strict monotonicity of $\ln(\cdot)$ follows easily that 
   \mbox{$f'(p_{\min}+\epsilon)>0>f'(p_{\min}-\epsilon)$} for all $\epsilon>0$, hence 
   $f$ attains a minimum at $p_{\min}$.
   The constraint $p_{\min}\geq 1$ is strictly fulfilled for every allowed $\alpha$ and $r$, hence
   the solution is feasible ($p_{\min}\geq\frac{2}{r}$) if $\alpha\geq\frac{2-r}{2}$. 
   Then the optimal (minimal) bound is
   \begin{equation}
      f(p_{\min})=\frac{2\alpha}{2\alpha+r}
   \end{equation}
   For the infeasible case instead, i.e. for $0<\alpha<\frac{2-r}{2}$, follows that minimal bound is attained 
   at the boundary point $p=\frac{2}{r}$. Thus $f(2/r)=\alpha^\frac{r}{2}(1-r/2)^{1-r/2}$.
   Summarizing,
   \begin{equation}
      \min_{\Reals\ni p\geq\max\{1,\frac{2}{r}\}}f(p)=
      \begin{cases}
         \frac{2\alpha}{2\alpha+r} & \alpha\geq\frac{2-r}{2}\\
         \alpha^\frac{r}{2}(1-r/2)^{1-r/2}         & \text{else}
      \end{cases}
   \end{equation}
   is the best possible upper bound.
   
   It remains to investigate the conditions for equality.
   Lieb's inequality is fulfilled with equality if strictly $p_{\min}>\frac{2}{r}$ and 
   $g,\gamma$ are matched Gaussians. In this case follows
   \begin{equation}
      \Amb_{g\gamma}(x)=e^{-\frac{\pi}{2}(ax_1^2+\frac{1}{\alpha}x_2^2)+\langle B,x\rangle+C}
   \end{equation}
   for some $B\in\Complexes^2$, $a,C\in\Complexes$ and $\Real{a}>0$, thus $\Amb_{g\gamma}(\cdot)$ is
   a two--dimensional Gaussian. Next, to have equality in (\ref{eq:hoelderapplied}) $p>1$ and
   equation (\ref{eq:hoelderequality:conditions}), which is in this case
   \begin{equation}
      \begin{split}
         |\Amb_{g\gamma}(x)|
         &=e^{\Real{-\frac{\pi}{2}(ax_1^2+\frac{1}{\alpha}x_2^2)+\langle B,x\rangle+C}} \\
         &=\lambda\alpha e^{-\frac{\alpha\pi}{r(p-1)}|x|^2}
         =\lambda|\BHScat(x)|^\frac{1}{r(p-1)}
      \end{split}
   \end{equation}
   have to be fulfilled for almost every $x$. 
   Thus, it follows that $\Real{B}=(0,0)$, $\lambda\alpha e^{-\Real{C}}=1$,
   $\Real{a}=1$ and -- most important -- again $p=\frac{2\alpha}{r}+1$.
   But, this is obviously also the minimum if $\alpha\geq\frac{2-r}{2}$, hence in this
   and only this case equality is achieved.
\end{proof}
\vspace*{.5em}
It is remarkable that the sharp ''if and only if'' conclusion for Gaussians 
holds now for  $\alpha\geq\frac{2-r}{2}$. Lieb's inequality alone needs
$\alpha>\frac{2-r}{2}$ but in conjuction with H\"older's inequality this is relaxed.
The results are illustrated in Fig.\ref{fig:ambbound:gaus}.
Furthermore note that for $r=2$ every $p_{\min}$ is feasible. 

This result is important for so called bosonic Gaussian quantum channels 
\cite{holevo:propquantum,hall:gaussiannoise}, i.e.
$\BHScat(\cdot)$ is a two--dimensional Gaussian.
In other words, according to (\ref{eq:quantum:fidelity}) and $\BHScat(\cdot)$ as in 
Corollary \ref{cor:ambbound:gaussian}, the solution of the Gaussian fidelity problem
\cite{arxiv:0409063,jung:isit05} is
\begin{equation}
   \begin{split}
      \max_{g,\gamma}\Trace\{\Pro_g A(\Pro_\gamma)\}
      &=\max_{\Trace{X}=1,X>0}\lambda_{\max}(A(X))\\
      &=\frac{\alpha}{\alpha+1}
   \end{split}
\end{equation}
with Gaussian $g$ and $\gamma$ as already found in \cite{arxiv:0409063} using a different approach. 
But now this states the strong proposition that maximum fidelity is achieved {\it only} by coherent states. 

In radar and sonar applications and also for wireless communications the following upper bound is important.
It is related to the case where scattering occurs with constant power in some region of 
phase space (in this context also called time--frequency plane).
For example in wireless communications typically only the maximal dispersions in time
and frequency (maximum delay spread and maximum Doppler spread) are assumed and/or known for some 
pulse shape optimization. Those situations are covered by the following result:
\begin{mycorollary}
   Let $U\subset\Reals^2$ a Borel set, $|U|<\infty$ 
   and $\BHScat(x)=\frac{1}{|U|}\chi_U(x)$ its $\Leb{1}$--normalized characteristic function.
   Then for each $p\geq\max\{1,\frac{2}{r}\}$ holds
   \begin{equation}
      \lVert |\Amb_{g\gamma}|^r \BHScat \rVert_1 <
      \left(\frac{2}{rp|U|}\right)^{\frac{1}{p}}
   \end{equation}
   It is not possible to achieve equality. The sharpest bound is 
   \begin{equation}
      \lVert |\Amb_{g\gamma}|^r \BHScat \rVert_1 <
      \begin{cases}
         e^{-\frac{r|U|}{2e}} & |U|\leq 2e/r^*\\
         \left(\frac{2}{r^*|U|}\right)^{r/r^*} & \text{else}
      \end{cases}
      \label{eq:ambbound:supp:best}
   \end{equation}
   where $r^*=\max\{r,2\}$.
\end{mycorollary}

\begin{figure}[h]
   \includegraphics[width=\linewidth]{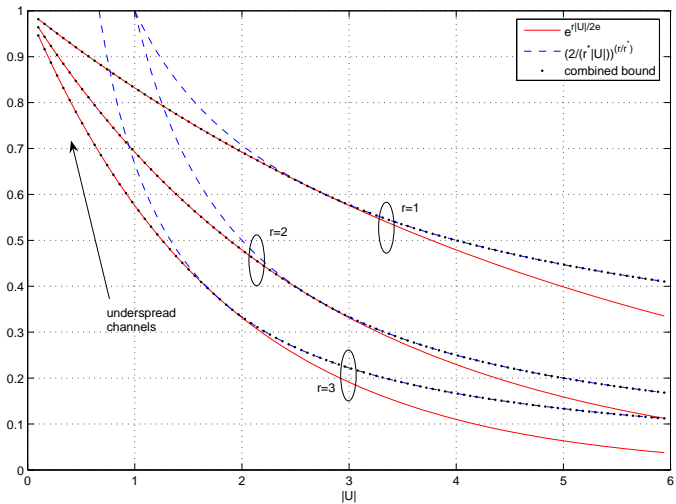}
   \caption{{\it Norm bounds for $\frac{1}{|U|}\chi_{U}$ weights:} Both functions in (\ref{eq:ambbound:supp:best})
     separately and the combined version are shown for $r=1,2,3$.}
   \label{fig:ambbound:supp}
\end{figure}

\begin{proof}
   The proof is straightforward by observing that 
   \begin{equation}
      \lVert\BHScat\rVert_s=\lVert\frac{1}{|U|}\chi_U\rVert_s=|U|^\frac{1-s}{s}
   \end{equation}
   According to Thm.\ref{thm:jung:fidelitybound} follows
   \begin{equation}
      \begin{split}
         f(p)
         &\defeq\left(\frac{2}{rp}\right)^{\frac{1}{p}}\lVert\BHScat\rVert_{\frac{p}{p-1}}
         ={\left(\frac{2}{rp|U|}\right)}^{\frac{1}{p}}=e^{-\frac{1}{p}\ln\frac{rp|U|}{2}}
      \end{split}
   \end{equation}

   Equality is not possible because Thm.\ref{thm:jung:fidelitybound} requires $\BHScat$ to be Gaussian 
   for equality.
   The optimal version is obtained by minimizing the function $f(p)$ under the constraint $p\geq\max\{1,2/r\}$.
   The first derivative $f'$ of $f$ at point $p$ is
   \begin{equation}
      f'(p)=\frac{f(p)}{p^2}(-\ln(\frac{2}{rp|U|})-1)
   \end{equation}
   Thus $f'(p_{\min})=0$ gives the only point $p_{\min}=\frac{2e}{r|U|}$. 
   The function $f(p)$ is log-convex on $p\in(0,\frac{2e^{3/2}}{r|U|}]\defeq I$. That 
   is $h(p)=\ln f(p)=-(\ln\frac{rp|U|}{2})/p$ is convex on $I$, because 
   \begin{equation}
      \begin{split}
         h'(p) &=\frac{1}{p^2}(\ln\frac{rp|U|}{2}-1) \\
         h''(p)&=\frac{1}{p^3}(-2\ln\frac{rp|U|}{2}+3)
      \end{split}
   \end{equation}
   shows, that $h''(p)\geq 0$ for all $p\in I$. Hence $f(p)$ is convex on $I$.
   Obviously 
   $p_{\min}\in I$, hence this point is in the convexity interval and therefore must be 
   the minimum of $f$. 
   Further, this value is also feasible if still $p_{\min}\geq\max\{1,2/r\}=r^*/r$
   where $r^*=\max\{r,2\}$, i.e. 
   \begin{equation}
      |U|< \frac{2e}{r^*}
   \end{equation}
   has to be
   fulfilled. Then the desired result is $f(p_{\min})=e^{-\frac{r|U|}{2e}}$.
   If $p_{\min}<r^*/r$, i.e. is infeasible, the
   minimum is attained at the boundary, i.e. at $p=r^*/r$. Thus 
   \begin{equation}
      f(r^*/r)=\left(\frac{2}{r^*|U|}\right)^{r/r^*}
   \end{equation}
\end{proof}
\vspace*{.5em}
The results are shown in Fig.\ref{fig:ambbound:supp} for $r=1,2,3$.
For the interesting case $r=2$ the result further simplifies to
\begin{equation}
   \lVert |\Amb_{g\gamma}|^2 \BHScat \rVert_1 <
   \begin{cases}
      e^{-\frac{|U|}{e}} & |U|\leq e\\
      |U|^{-1}     & \text{else}
   \end{cases}
\end{equation}
\emph{Example:} When using the WSSUS model \cite{bello:wssus} for doubly--dispersive 
mobile communication channels one typically assumes 
time--frequency scattering with shape
\begin{equation}
   U=\{(x_1,x_2)\,|\, 0\leq x_1\leq\tau_d\,,\, |x_2|\leq B_d\}
\end{equation}
with $2B_d\tau_d\ll 1< e$, where $B_d$ denotes maximum Doppler bandwidth $B_d$
and $\tau_d$ is maximum delay spread.
Then \eqref{eq:ambbound:supp:best} predicts, that 
the best (mean) correlation response ($r=2$) in using filter $g$ at the 
receiver and $\gamma$ at the transmitter is bounded above by 
\begin{equation}
   \lVert |\Amb_{g\gamma}|^2 \BHScat \rVert_1 < e^{-\frac{2B_d\tau_d}{e}}
\end{equation} 

\section{Conclusions}
In this contribution new bounds on weighted norms of ambiguity functions and
Wigner distributions are presented which only depend on the shape of the weight function.
Further the important equality case is discussed which is attained only by Gaussian weights
and wave functions. The results are important in the field of waveform optimization for
non--stationary environments as needed for example in WSSUS channels. This channel model is
frequently used in radar and sonar applications and -- of course -- in wireless communications.
Furthermore these norms are needed in quantum information processing for bosonic quantum channels
because they provide insights on achievable fidelities in those quantum channels.
In the special case of the Gaussian quantum channel they provide also the optimum input states, 
i.e. {\it only} coherent states achieve this optimal fidelity as frequently conjectured.
Hence, in the mentioned fields the results establish limits on achievable performance.

\section{Acknowledgments}
I would like to thank Igor Bjelakovic for many useful discussions.

\bibliographystyle{IEEEtran}
\bibliography{references}
\end{document}